\documentstyle[11pt,epsfig]{article}
\title{\bf Brane-World Black Hole Solutions via a Confining Potential}
\author{M. Heydari-Fard $^{1}$\thanks{email:
 m.heydarifard@mail.sbu.ac.ir},
H. Razmi $^{2}$\thanks{email: razmi@qom.ac.ir } and H. R. Sepangi
$^{1}$\thanks{email: hr-sepangi@sbu.ac.ir}
\\ {\small Department
of Physics, Shahid Beheshti University, Evin, Tehran 19839,
Iran}$^{1}$
\\ {\small Department
of Physics, The University of Qom, Qom 37185-359, Iran}$^{2}$}
\begin{document}
\maketitle %\baselineskip 24pt
\begin{abstract}
Using a confining potential, we consider spherically symmetric
vacuum (static black hole) solutions in a brane-world scenario.
Working with a constant curvature bulk, two interesting
cases/solutions are studied. A Schwarzschild-de Sitter black hole
solution similar to the standard solution in the presence of a
cosmological constant is obtained which confirms the idea that an
extra term in the field equations on the brane can play the role
of a positive cosmological constant and may be used to account for
the accelerated expansion of the universe. The other solution is
one in which we can have a proper potential to explain the galaxy
rotation curves without assuming the existence of dark matter and
without working with new modified theories (modified Newtonian
dynamics).
\vspace{5mm}
\end{abstract}
\section{Introduction}
In recent years, models with extra dimensions have been proposed
in which the standard fields are confined to a four-dimensional
($4D$) world viewed as a hypersurface (the brane) embedded in a
higher dimensional space-time (the bulk) through which only
gravity can propagate. The most well-known model in the context of
brane-world theory is that proposed by Randall and Sundrum (RS).
In the so-called RSI model \cite{11}, they proposed a mechanism to
solve the hierarchy problem with two branes, while in the RSII
model \cite{12}, they considered a single brane with a positive
tension, where $4D$ Newtonian gravity is recovered at low energies
even if the extra dimension is not compact. This mechanism
provides an alternative to compactification of extra dimensions.

The cosmological evolution of such a brane universe has been
extensively investigated and effects such as a quadratic density
term in the Friedmann equations have been found \cite{13,14}. This
term arises from the imposition of the Israel junction conditions
which is a relationship between the extrinsic curvature and
energy-momentum tensor of the brane and that results from the
singular behavior in the energy-momentum tensor. There has been
concerns expressed over applying such junction conditions in that
they may not be unique. Indeed, other forms of junction conditions
exist, so that different conditions may lead to different physical
results \cite{16}. Furthermore, these conditions cannot be used
when more than one non-compact extra dimension is involved. To
avoid such concerns, an interesting higher-dimensional model was
introduced where particles are trapped on a 4-dimensional
hypersurface by the action of a confining potential ${\cal V}
$\cite{17} . In \cite{18}, the dynamics of test particles confined
to a brane by the action of such a potential at the classical and
quantum levels were studied and the effects of small perturbations
along the extra dimensions investigated. In \cite{19}, a
brane-world model was studied in which matter is confined to the
brane through the action of such a potential without using any
junction conditions, offering a geometrical explanation for the
accelerated expansion of the universe. Another work in where
localization of matter on the brane is again realized by means of
a confining potential is the study of a brane scenario in which
the $m$-dimensional bulk is endowed with a Gauss-Bonnet (GB) term
\cite{Gauss-Bonnet}. It was shown that in the presence of the GB
term, the universe accelerates faster than brane models without
the GB term. The behavior of an anisotropic brane-world with
Bianchi type I and V geometry in a similar vain was studied in
\cite{anisotropic}.

In brane theories the covariant Einstein equations are derived by
projecting the bulk equations onto the brane. This was first done
by Shiromizu, Maeda and Sasaki (SMS) [11] where the Gauss-Codazzi
equations together with Israel junction conditions were used to
obtain the Einstein field equations on the 3-brane. The field
equations on the brane is different from the Einstein equations in
the standard model. An essential modification appears at high
energies in the form of a new source term in the effective
Einstein equation, which is quadratic in the brane energy-momentum
tensor. Another modification arises whenever the bulk has a
Weyl-curvature with non-vanishing projection onto the brane. This
is known as the electric part of the bulk Weyl tensor. In this
context, it is natural to study solutions corresponding to compact
sources on the brane such as stars and black holes. Gravitational
collapse on the brane has been studied by many authors
\cite{Collapse}-\cite{collapse}. In \cite{naresh}, the authors
obtain an exact black hole solution of the effective Einstein
equation on the brane under the condition that the bulk has non
zero Weyl curvature and the brane space-time satisfies the null
energy condition. The solution is given by the usual
Reissner-Nordstrom (RN) metric where the charge parameter is
thought of as a tidal charge arising from the projection of the
Weyl curvature of the bulk onto the brane. The RN metric has thus
been interpreted as describing a black hole on a brane where the
electric charge's role is taken over by the tidal charge and it
can be thought of as the analogue of the Schwarzschild solution on
the brane. The tidal charge like the RN electric charge would
generate $1/r^2$ term in the potential while the high energy
modification to the Newtonian potential cannot be any stronger
than $1/r^3$ \cite{12,New}. The cause for this disagreement is the
presence of tidal charge which is the measure of the bulk Weyl
curvature. The main drawback of the solution is that we do not
know the corresponding bulk solution. It is however agreed that RN
metric is a good approximation to a black hole on the brane near
the horizon \cite{new}.

The RN solution can be matched to the interior solution
corresponding to a constant density brane-world star. A second
exterior solution, which also matches a constant density interior,
has been derived in \cite{22}. Non-singular black hole solutions
in the brane-world model have been considered in \cite{21}, by
relaxing the condition of the zero scalar curvature but retaining
the null energy condition. It has also been shown that the vacuum
field equations on the brane reduce to a system of two ordinary
differential equations, which describe all the geometric
properties of the vacuum as functions of the dark pressure and
dark radiation terms \cite{mak}.

In this paper, following the model introduced in \cite{19}, we
consider an $m$-dimensional bulk space without imposing the $Z_2$
symmetry. As mentioned above, to localize the matter on the brane,
a confining potential is used rather than a delta-function in the
energy-momentum tensor. The vacuum field equations on the brane
are modified by the $Q_{\mu\nu}$ term which is a geometrical
quantity. We obtain exact solutions of the vacuum field equations
on the brane for two interesting cases. The first solution can be
used to explain the galaxy rotation curves without assuming the
existence of dark matter and without working with new modified
theories \cite{3} (modified Newtonian dynamics) and the second
solution represents a black hole in an asymptotically de Sitter
space. Clearly, this work differs from the model introduced in
\cite{23,24} in them no mechanism for the confinement of matter on
the brane is introduced.
\section{Geometrical considerations}
In this section we present a brief review of the model proposed in
\cite{18,19}.  Consider the background manifold $ \overline{V}_{4}
$ isometrically embedded in a pseudo-Riemannian manifold $ V_{m}$
by the map ${ \cal Y}: \overline{V}_{4}\rightarrow  V_{m} $ such
that
\begin{eqnarray}\label{a}
{\cal G} _{AB} {\cal Y}^{A}_{,\mu } {\cal Y}^{B}_{,\nu}=
\bar{g}_{\mu \nu}  , \hspace{.5 cm} {\cal G}_{AB}{\cal
Y}^{A}_{,\mu}{\cal N}^{B}_{a} = 0  ,\hspace{.5 cm}  {\cal
G}_{AB}{\cal N}^{A}_{a}{\cal N}^{B}_{b} = \bar{g}_{ab}= \pm 1,
\end{eqnarray}
where $ {\cal G}_{AB} $  $ ( \bar{g}_{\mu\nu} ) $ is the metric of
the bulk (brane) space  $  V_{m}  (\overline{V}_{4}) $ in
arbitrary coordinates, $ \{ {\cal Y}^{A} \} $   $  (\{ x^{\mu} \})
$  is the  basis of the bulk (brane) and  ${\cal N}^{A}_{a}$ are
$(m-4)$ normal unit vectors, orthogonal to the brane. Perturbation
of $\bar{V}_{4}$ in a sufficiently small neighborhood of the brane
along an arbitrary transverse direction $\xi$ is given by
\begin{eqnarray}\label{a1}
{\cal Z}^{A}(x^{\mu},\xi^{a}) = {\cal Y}^{A} + ({\cal
L}_{\xi}{\cal Y})^{A}, \label{eq2}
\end{eqnarray}
where $\cal L$ represents the Lie derivative and $\xi^{a}$ $(a =
1,2,...,m-4)$ is a small parameter along ${\cal N}^{A}_{a}$ that
parameterizes the extra noncompact dimensions. By choosing $\xi$
orthogonal to the brane, we ensure gauge independency \cite{18}
and have perturbations of the embedding along a single orthogonal
extra direction $\bar{{\cal N}}_{a}$ giving local coordinates of
the perturbed brane as
\begin{eqnarray}\label{a2}
{\cal Z}^{A}_{,\mu}(x^{\nu},\xi^{a}) = {\cal Y}^{A}_{,\mu} +
\xi^{a}\bar{{\cal N}}^{A}_{a,\mu}(x^{\nu}).
\end{eqnarray}
In a similar manner, one can find that since the vectors
$\bar{{\cal N}}^{A}$ depend only on the local coordinates
$x^{\mu}$, they do not propagate along the extra dimensions. The
above  assumptions lead to the embedding equations of the
perturbed geometry
\begin{eqnarray}\label{a4}
g_{\mu \nu }={\cal G}_{AB}{\cal Z}_{\,\,\ ,\mu }^{A}{\cal
Z}_{\,\,\ ,\nu }^{B},\hspace{0.5cm}g_{\mu a}={\cal G}_{AB}{\cal
Z}_{\,\,\ ,\mu }^{A}{\cal N}_{\,\,\ a}^{B},\hspace{0.5cm}{\cal
G}_{AB}{\cal N}_{\,\,\
a}^{A}%
{\cal N}_{\,\,\ b}^{B}={g}_{ab}.
\end{eqnarray}
If we set ${\cal N}_{\,\,\ a}^{A}=\delta _{a}^{A}$, the metric of
the bulk space can be written in the following matrix form
\begin{eqnarray}
{\cal G}_{AB}=\left( \!\!\!%
\begin{array}{cc}
g_{\mu \nu }+A_{\mu c}A_{\,\,\nu }^{c} & A_{\mu a} \\
A_{\nu b} & g_{ab}%
\end{array}%
\!\!\!\right) ,  \label{F}
\end{eqnarray}%
where
\begin{eqnarray}
g_{\mu \nu }=\bar{g}_{\mu \nu }-2\xi ^{a}\bar{K}_{\mu \nu a}+\xi
^{a}\xi ^{b}%
\bar{g}^{\alpha \beta }\bar{K}_{\mu \alpha a}\bar{K}_{\nu \beta
b}, \label{G}
\end{eqnarray}%
is the metric of the perturbed brane, so that
\begin{eqnarray}
\bar{K}_{\mu \nu a}=-{\cal G}_{AB}{\cal Y}_{\,\,\,,\mu }^{A}{\cal
N}_{\,\,\ a;\nu }^{B},  \label{H}
\end{eqnarray}%
represents the extrinsic curvature of the original brane (second
fundamental form). We use the notation $A_{\mu c}=\xi ^{d}A_{\mu
cd}$, where
\begin{equation}
A_{\mu cd}={\cal G}_{AB}{\cal N}_{\,\,\ d;\mu }^{A}{\cal N}_{\,\,\
c}^{B}=%
\bar{A}_{\mu cd},  \label{I}
\end{equation}%
represents the twisting vector fields
(the normal fundamental form). Any fixed $%
\xi ^{a}$ signifies a new perturbed geometry, enabling us to
define an extrinsic curvature similar to the original one by
\begin{eqnarray}
\widetilde{K}_{\mu \nu a}=-{\cal G}_{AB}{\cal Z}_{\,\,\ ,\mu
}^{A}{\cal
N}%
_{\,\,\ a;\nu }^{B}=\bar{K}_{\mu \nu a}-\xi ^{b}\left(
\bar{K}_{\mu
\gamma a}%
\bar{K}_{\,\,\ \nu b}^{\gamma }+A_{\mu ca}A_{\,\,\ b\nu
}^{c}\right) . \label{J}
\end{eqnarray}%
Note that definitions (\ref{F}) and (\ref{J}) require
\begin{eqnarray}
\widetilde{K}_{\mu \nu a}=-\frac{1}{2}\frac{\partial {\cal G}_{\mu
\nu
}}{%
\partial \xi ^{a}}.  \label{M}
\end{eqnarray}%
In geometric language, the presence of gauge fields $A_{\mu a}$
tilts the embedded family of sub-manifolds with respect to the
normal vector ${\cal N} ^{A}$. According to our construction, the
original brane is orthogonal to the normal vector ${\cal N}^{A}.$
However,  equation (\ref{a4})  shows that this is not true for the
deformed geometry. Let us change the embedding coordinates and set
\begin{eqnarray}
{\cal X}_{,\mu }^{A}={\cal Z}_{,\mu }^{A}-g^{ab}{\cal
N}_{a}^{A}A_{b\mu }. \label{mama40}
\end{eqnarray}%
The coordinates ${\cal X}^{A}$ describe a new family of embedded
manifolds whose members are always orthogonal to ${\cal N}^{A}$.
In this coordinates the embedding equations of the perturbed brane
is similar to the original one, described by equation (\ref{a}),
so that ${\cal Y}^{A}$ is replaced by ${\cal X}^{A}$. This new
embedding of the local coordinates are suitable for obtaining
induced Einstein field equations on the brane. The extrinsic
curvature of a perturbed brane then becomes
\begin{eqnarray}
K_{\mu \nu a}=-{\cal G}_{AB}{\cal X}_{,\mu }^{A}{\cal N}_{a;\nu
}^{B}=\bar{K}%
_{\mu \nu a}-\xi ^{b}\bar{K}_{\mu \gamma a}\bar{K}_{\,\,\nu
b}^{\gamma
}=-%
\frac{1}{2}\frac{\partial g_{\mu \nu }}{\partial \xi ^{a}},
\label{mama42}
\end{eqnarray}%
which is the generalized York's relation and shows how the
extrinsic curvature propagates as a result of the propagation of
the metric in the direction of extra dimensions. The components of
the Riemann tensor of the bulk written in the embedding vielbein
$\{{\cal X}^{A}_{, \alpha}, {\cal N}^A_a \}$, lead to the
Gauss-Codazzi equations \cite{27}
\begin{eqnarray}\label{a5}
R_{\alpha \beta \gamma \delta}=2g^{ab}K_{\alpha[ \gamma
a}K_{\delta] \beta b}+{\cal R}_{ABCD}{\cal X} ^{A}_{,\alpha}{\cal
X} ^{B}_{,\beta}{\cal X} ^{C}_{,\gamma} {\cal X}^{D}_{,\delta},
\end{eqnarray}
\begin{eqnarray}\label{a6}
2K_{\alpha [\gamma c; \delta]}=2g^{ab}A_{[\gamma ac}K_{ \delta]
\alpha b}+{\cal R}_{ABCD}{\cal X} ^{A}_{,\alpha} {\cal N}^{B}_{c}
{\cal X} ^{C}_{,\gamma} {\cal X}^{D}_{,\delta},
\end{eqnarray}
where ${\cal R}_{ABCD}$ and $R_{\alpha\beta\gamma\delta}$ are the
Riemann tensors for the bulk and the perturbed brane respectively.
Contracting the Gauss equation (\ref{a5}) on ${\alpha}$ and
${\gamma}$, we find
\begin{eqnarray}\label{a7}
R_{\mu\nu}=(K_{\mu\alpha c}K_{\nu}^{\,\,\,\,\alpha c}-K_{c} K_{\mu
\nu }^{\,\,\,\ c})+{\cal R}_{AB} {\cal X}^{A}_{,\mu} {\cal
X}^{B}_{,\nu}-g^{ab}{\cal R}_{ABCD}{\cal N}^{A}_{a}{\cal
X}^{B}_{,\mu}{\cal X}^{C}_{,\nu}{\cal N}^{D}_{b}.
\end{eqnarray}
A further contraction gives the Ricci scalar
\begin{eqnarray}\label{a7new}
R=(K_{\mu\alpha c}K^{\mu\alpha c}-K_cK^c)+{\cal R}-2g^{ab}{\cal
R}_{AB}{\cal N}^{A}_{a}{\cal N}^{B}_{b}+g^{ad}g^{bc}{\cal
R}_{ABCD}{\cal N}^{A}_{a}{\cal N}^{B}_{b}{\cal N}^{C}_{c}{\cal
N}^{D}_{d}.
\end{eqnarray}
\section{Field equations on the brane}
We consider the total action for space-time $({\cal M},{\cal
G}_{AB})$ with boundary $(\Sigma,g_{\mu\nu})$ as
\begin{eqnarray}\label{mm}
S=\frac{1}{2\alpha^{*}}\int_{\cal M} d^{m}X \sqrt{-\cal G} ({\cal
R}-2\Lambda^{(b)})+\int_{\Sigma} d^{4}x \sqrt{-g} ({\cal L}_{\rm
surface}+{\cal L}_{m}).
\end{eqnarray}
Variation of the total action gives the Einstein equations in the
bulk space as
\begin{eqnarray}
{G}^{(b)}_{AB} +\Lambda^{(b)}{\cal G}_{AB}=\alpha^{*}
S_{AB},\label{eqq14}
\end{eqnarray}
where
\begin{eqnarray}\label{a14}
S_{AB}=T_{AB}+ \frac{1}{2} {\cal V} {\cal G}_{AB},\label{eqq15}
\end{eqnarray}
here $\alpha^{*}=\frac{1}{M_{*}^{m-2}}$ ($M_{*}$ is the
fundamental scale of energy in the bulk space), $\Lambda^{(b)}$ is
the cosmological constant of the bulk and $T_{AB}\equiv -2 \frac
{\delta {\cal L}_{\rm m} }{ \delta g^{AB}} +g_{AB}{\cal L}_{\rm
m}$ is the energy-momentum tensor of the matter confined to the
brane through the action of the confining potential $\cal{V}$. We
require $\cal{V}$  to satisfy three general conditions: firstly,
it has a deep minimum on the non-perturbed brane, secondly,
depends only on extra coordinates and thirdly, the gauge group
representing the subgroup of the isometry group of the bulk space
is preserved by it \cite{18}.

Now, using equations (\ref{a7}) and (\ref{a7new}) the Einstein
tensor on the brane is given by
\begin{eqnarray}\label{a8}
G_{\mu\nu}=G_{AB} {\cal X}^{A}_{,\mu}{\cal
X}^{B}_{,\nu}+Q_{\mu\nu}+g^{ab}{\cal R}_{AB}{\cal N}^{A}_{a}{\cal
N}^{B}_{b} g_{\mu\nu}- g^{ab}{\cal R}_{ABCD}{\cal N}^{A}_{a}{\cal
X}^{B}_{,\mu}{\cal X}^{C}_{,\nu}{\cal N}^{D}_{b},
\end{eqnarray}
where
\begin{eqnarray}\label{a9}
Q_{\mu\nu}=-g^{ab}\left(K^\gamma_{\mu a}K_{\gamma\nu b}-K_a
K_{\mu\nu b}\right)+\frac{1}{2}\left(K_{\alpha\beta
a}K^{\alpha\beta a}-K_a K^a\right)g_{\mu\nu}. \label{eqq7}
\end{eqnarray}
As can be seen from the definition of $Q_{\mu\nu}$,  it is an
independently conserved quantity, that is $Q^{\mu\nu}_{\,\,\,\,\,
;\nu}=0$ \cite{23}. Using the decomposition of the Riemann tensor
into the Weyl curvature, the Ricci tensor and the scalar curvature
\begin{eqnarray}\label{a10}
{\cal R}_{ABCD}=C_{ABCD}-\frac{2}{(m-2)}\left({\cal G}_{B[D}{\cal
R}_{C]A}-{\cal G}_{A[D}{\cal
R}_{C]B}\right)-\frac{2}{(m-1)(m-2)}{\cal R}({\cal G}_{A[D}{\cal
G}_{C]B}),
\end{eqnarray}
we obtain the $4D$ Einstein equations as
\begin{eqnarray}\label{a11} G_{\mu\nu}&=&G_{AB} {\cal
X}^{A}_{,\mu}{\cal X}^{B}_{,\nu}+Q_{\mu\nu}-{\cal
E}_{\mu\nu}+\frac{m-3}{(m-2)}g^{ab}{\cal R}_{AB}{\cal
N}^{A}_{a}{\cal
N}^{B}_{b}g_{\mu\nu}\nonumber\\
&-&\frac{m-4}{(m-2)}{\cal R}_{AB}{\cal X}^{A}_{,\mu}{\cal
X}^{B}_{,\nu}+\frac{m-4}{(m-1)(m-2)}{\cal
R}g_{\mu\nu},\label{eqq12}
\end{eqnarray}
where ${\cal E}_{\mu\nu}=g^{ab} C_{ABCD}{\cal N}^{A}_{a}{\cal
X}^{B}_{,\mu}{\cal N}^D_b{\cal X}^C_{,\nu}$ is the electric part
of the Weyl tensor of the bulk space $C_{ABCD}$. Using the
Einstein equations (\ref{eqq14}), we obtain
\begin{eqnarray}\label{a15}
{\cal R}_{AB}=-\frac{\alpha^{*}}{(m-2)}{\cal G}_{AB}
S+\frac{2}{(m-2)}\Lambda^{(b)} {\cal G}_{AB}+\alpha^{*} S_{AB},
\end{eqnarray}
and
\begin{eqnarray}\label{a16}
{\cal R}=-\frac{2}{m-2}(\alpha^{*} S-m\Lambda^{(b)}).
\end{eqnarray}
Substituting ${\cal R}_{AB}$ and ${\cal R}$ from the above into
equation (\ref{a11}), we obtain
\begin{eqnarray}\label{a17}
G_{\mu\nu}&=& Q_{ \mu\nu} - {\cal
E}_{\mu\nu}+\frac{(m-3)}{(m-2)}\alpha^{*}g^{ab}S_{ab}g_{\mu\nu}
+\frac{2\alpha^{*}}{(m-2)}S_{\mu\nu} -
\frac{(m-4)(m-3)}{(m-1)(m-2)}\alpha^{*}Sg_{\mu\nu}\nonumber\\
&+&\frac{(m-7)}{(m-1)}\Lambda^{(b)}g_{\mu\nu}.\label{new1}
\end{eqnarray}
On the other hand, again from equation (\ref{eqq14}), the trace of
the Codazzi equation (\ref{a6}) gives the ``gravi-vector
equation''
\begin{equation}
K^\delta_{a\gamma;\delta} - K_{a,\gamma} - A_{ba\gamma}K^b +
A_{ba\delta}K^{b\delta} + B_{a\gamma} =
\frac{3(m-4)}{m-2}\alpha^{*}S_{a\gamma},\label{new2}
\end{equation}
where
\begin{equation}
B_{a\gamma} = g^{mn}C_{ABCD}{\cal N}^A_m{\cal N}^B_a{\cal
X}^C_{,\gamma}{\cal N}^D_n.
\end{equation}
Finally, the ``gravi-scalar equation'' is obtained from the
contraction of (\ref{a7}), (\ref{a11}) and using equation
(\ref{eqq14})
\begin{equation}
\alpha^{*}\left[\frac{m-5}{m-1}S - g^{mn}S_{mn}\right]g_{ab} =
\frac{m-2}{6}\left(Q + R +W\right)g_{ab} -
\frac{4}{m-1}\Lambda^{(b)}g_{ab},\label{new3}
\end{equation}
where
\begin{equation}
W = g^{ab}g^{mn}C_{ABCD}{\cal N}^A_m{\cal N}^B_b{\cal N}^C_n{\cal
N}^D_a.\label{New3}
\end{equation}
Equations (\ref{new1})-(\ref{New3}) represent the projections of
the Einstein field equations on the brane-brane, bulk-brane, and
bulk-bulk directions.

As was mentioned in the introduction, localization of matter on
the brane is realized in this model by the action of a confining
potential. This can simply be realized by
\begin{eqnarray}
\alpha\tau_{\mu\nu} = \frac{2\alpha^{*}}{(m-2)}T_{\mu\nu},
\hspace{.5 cm}T_{\mu a}=0, \hspace{.5 cm}T_{ab}=0,\label{new4}
\end{eqnarray}
where $\alpha$ is the scale of energy on the brane. Now, the
induced Einstein field equations on the original brane can be
written as
\begin{eqnarray}
G_{\mu\nu} = \alpha \tau_{\mu\nu}
-\frac{(m-4)(m-3)}{2(m-1)}\alpha\tau g_{\mu\nu} - \Lambda
g_{\mu\nu} + Q_{\mu\nu} - {\cal E}_{\mu\nu},\label{a8}
\end{eqnarray}
where  $\Lambda= -\frac{(m-7)}{(m-1)} \Lambda^{(b)}$ and
$Q_{\mu\nu}$ is a completely geometrical quantity.

A brief discussion on the energy-momentum conservation on the
brane would be in order here. The contracted Bianchi identities in
the bulk space $G^{AB(b)}_{\,\,\,\,\,\,\,\,\,;A}=0$, using
equation (\ref{eqq14}), imply
\begin{eqnarray}
\left(T^{AB}+\frac{1}{2} {\cal{V}} {\cal {G}}^{AB}\right)_{
;A}=0.\label{eq1}
\end{eqnarray}
Since the potential $\cal V$ has a minimum on the brane, the above
conservation equation reduces to
\begin{eqnarray}\label{a22}
\tau^{\mu\nu}_{\,\,\,\,\,;\mu}=0.
\end{eqnarray}
As we mentioned before, $Q_{\mu\nu}$ is an independently conserved
quantity which according to \cite{23} may be considered as an
energy-momentum tensor of a dark energy fluid representing the
x-matter, the more common phrase being ``x-Cold-Dark Matter''
(xCDM). This matter has the most general form of the equation of
state which is characterized by the following conditions
\cite{25}:  violation of the strong energy condition at the
present epoch for $\omega_x<-1/3$ where $p_x=\omega_x\rho_x$,
local stability {\it i.e.} $c^2_s=\delta p_x/\delta\rho_x\ge 0$
and preservation of causality {\it i.e.} $c_s\le 1$. Ultimately,
we have three different types of `matter' on the right hand side
of equation (\ref{a8}), namely, ordinary confined conserved matter
represented by $\tau_{\mu\nu}$, the matter represented by
$Q_{\mu\nu}$, the Weyl matter represented by ${\cal E}_{\mu\nu}$.
\section{Vacuum solutions on the brane}
For the vacuum field equations with $\tau_{\mu\nu}=0$ and thus
$\tau=0$, the effective equations derived in the previous section
are given by
\begin{eqnarray}
G_{\mu\nu}=- \Lambda g_{\mu\nu} + Q_{\mu\nu} - {\cal
E}_{\mu\nu},\label{A9}
\end{eqnarray}
where ${\cal E}_{\mu\nu}$ is a symmetric and traceless tensor due
to the Weyl symmetries and is constrained by the conservation
equations
\begin{eqnarray}
{\cal E}^{\mu\nu}_{\,\,\,\,\ ;\nu}=0,\label{A10}
\end{eqnarray}
obtained as a result of the Bianchi identities. Equations
(\ref{A9}) and (\ref{A10}) determine the system of vacuum field
equations on the brane. Restricting our analysis to a constant
curvature bulk (${\cal E}_{\mu\nu}=0$) and neglecting the effect
of the cosmological constant, the vacuum field equation (\ref{A9})
reduce to
\begin{eqnarray}
G_{\mu\nu} = Q_{\mu\nu},\label{1}
\end{eqnarray}
where $Q_{\mu\nu}$ is a completely geometrical quantity given by
\begin{eqnarray}\label{new5}
Q_{\mu\nu}=\left(KK_{\mu\nu}- K_{\mu\alpha
}K^{\alpha}_{\nu}\right)+\frac{1}{2}
\left(K_{\alpha\beta}K^{\alpha\beta}-K^2\right)g_{\mu\nu}.\label{new5}
\end{eqnarray}
For the following choice of the static spherically symmetric
metric on the brane
\begin{eqnarray}
ds^2=-e^{\mu(r)}dt^2+e^{\nu(r)}dr^2+r^2(d\theta^2+sin^2\theta
d\varphi^2),\label{2}
\end{eqnarray}
the gravitational field equations are given by
\begin{eqnarray}
G_{0}^{0}=\frac{e^{-\nu}}{r^2}\left(1-r\nu^{'}-e^{\nu}\right),\label{3}
\end{eqnarray}
\begin{eqnarray}
G_{1}^{1}=\frac{e^{-\nu}}{r^2}\left(1+r\mu^{'}-e^{\nu}\right),\label{4}
\end{eqnarray}
\begin{eqnarray}
G_{2}^{2}=G_{3}^{3}=\frac{e^{-\nu}}{4r}\left(2\mu^{'}-2\nu^{'}-\mu^{'}\nu^{'}
r+2\mu^{''}r+\mu^{'2}r\right),\label{new4}
\end{eqnarray}
where a prime represents differentiation with respect to $r$. The
York's relation
\begin{eqnarray}
K_{\mu \nu a}=-\frac{1}{2}\frac{\partial
g_{\mu\nu}}{\partial\xi^{a}},\label{5}
\end{eqnarray}
shows that in a diagonal metric, $K_{\mu\nu a}$ are diagonal. The
Codazzi equations (\ref{a6}) with the assumption of ${\cal
E_{\mu\nu}}=0$ take the form
\begin{eqnarray}
K_{\alpha\delta a;\gamma}-K_{\alpha\gamma a;\delta}=0,\label{6}
\end{eqnarray}
where by separating the spatial components reduce to
\begin{equation}\label{a28}
K_{\mu\nu a,\sigma}-K_{\nu\rho a}\Gamma^{\rho}_{\mu\sigma}=
K_{\mu\sigma a,\nu}-K_{\sigma\rho
a}\Gamma^{\rho}_{\mu\nu},\label{7}
\end{equation}
\begin{eqnarray}
K_{00a,1}-\left(\frac{\mu{'}}{2}\right)K_{00a}=-\left(\frac{\mu{'}e^{\mu}}{2e^{\nu}}\right)
K_{11a},\label{8}
\end{eqnarray}
\begin{eqnarray}
K_{22a,1}-\left(\frac{1}{r}\right)K_{22a}=\left(re^{-\nu}\right)
K_{11a}.\label{9}
\end{eqnarray}
The first equation gives
$K_{00a,\sigma}=K_{11a,\sigma}=K_{22a,\sigma}=K_{33a,\sigma}=0$
for $\sigma=0,3$. Repeating the same procedure for $\sigma=2$, we
obtain $K_{00a,\sigma}=K_{11a,\sigma}=K_{22a,\sigma}=0$. This
shows that $K_{11a}$ depends only on the variable $r$. Assuming
$K_{11a}=\alpha_{a}e^{\nu(r)}$ and using equations (\ref{8}) and
(\ref{9}), one finds
\begin{eqnarray}
K_{00a}(r)=-\alpha_{a}e^{\mu(r)} +c_{a}e^{\mu(r)/2},\label{10}
\end{eqnarray}
\begin{eqnarray}
K_{22a}(r)=\alpha_{a} r^2+\beta_{a} r.\label{11}
\end{eqnarray}
Taking $\mu,\nu=3$ in the first equation we obtain
\begin{eqnarray}
K_{33a,1}-\left(\frac{1}{r}\right)K_{33a}=\left(e^{-\nu}
r\sin^2{\theta}\right)K_{11a}=\alpha_{a}r\sin^2{\theta},\label{12}
\end{eqnarray}
\begin{eqnarray}
K_{33a,2}-\left(\cot{\theta}\right)K_{33a}=\left
(\sin{\theta}\cos{\theta}\right)K_{22a}.\label{13}
\end{eqnarray}
Using equations (\ref{11}), (\ref{12}) and (\ref{13}), $k_{33a}$
is given by
\begin{eqnarray}
K_{33a}(r,\theta)=\alpha_{a} r^2\sin^2{\theta}+r\beta_{a}
\sin^2{\theta}+rc_{1a}\sin{\theta}.\label{14}
\end{eqnarray}
If we assume that the  constants are equal, that is
$\alpha_{a}=\alpha$, $\beta_{a}=\beta$ and $ c_{a}=c,$
$c_{1a}=c_{1}$, use of equation (\ref{new5}) leads to the
components of $Q_{\mu\nu}$
\begin{eqnarray}\label{EQ}
Q_{00}&=&-\frac{g_{00}}{r^2} \left(3\alpha^2r^2+4\alpha\beta
r+\beta^2+\frac{c_1}{\sin{\theta}}(2\alpha
r+\beta) \right),\nonumber\\
Q_{11}&=&- \frac{g_{11}}{r^2}\left(3\alpha^2r^2+4\alpha\beta
r+\beta^2+\frac{c_1}{\sin{\theta}}(2\alpha r+\beta-c
re^{-\mu/2})-2ce^{-\mu/2}(\alpha r^2+\beta
r)\right),\nonumber\\
Q_{22}&=&
\frac{g_{22}}{r}\left(-3\alpha^2r-2\alpha\beta+ce^{-\mu/2}(2\alpha
r+\beta)+\frac{c_1}{\sin{\theta}}
(-2\alpha+ce^{-\mu/2})\right),\nonumber\\
Q_{33}&=&
\frac{g_{33}}{r}\left(-3\alpha^2r-2\alpha\beta+ce^{-\mu/2}(2\alpha
r+\beta)\right).
\end{eqnarray}
Since $G_{2}^{2}=G_{3}^{3}$ and thus $Q_{2}^{2}=Q_{3}^{3}$ one
obtains $c_1=0$. The relations (\ref{EQ}), equations (\ref{1}) and
(\ref{3})-(\ref{new4}) lead to the vacuum field equations on the
brane
\begin{eqnarray}\label{17}
{e^{-\nu}}\left(-\frac{1}{r^2}+\frac{\nu^{'}}{r}\right)+\frac{1}{r^2}=3\alpha^2+\frac{4\alpha\beta}{r}
+\frac{\beta^2}{r^2},\label{17}
\end{eqnarray}
\begin{eqnarray}\label{18}
-e^{-\nu}\left(\frac{1}{r^2}+\frac{\mu^{'}}{r}\right)+\frac{1}{r^2}=3\alpha^2+\frac{4\alpha\beta}{r}
+\frac{\beta^2}{r^2}-2ce^{-\mu/2}(\alpha+\frac{\beta}{r}
),\label{18}
\end{eqnarray}
\begin{eqnarray}\label{new18}
e^{-\nu}\left(\frac{\mu^{'}-\nu^{'}}{r}-\frac{\mu^{'}\nu^{'}}{2}
+\mu^{''}+\frac{\mu^{'2}}{2}\right)=-6\alpha^2-\frac{4\alpha\beta}{r}+2ce^{-\mu/2}(2\alpha
+\frac{\beta}{r}).\label{new18}
\end{eqnarray}
Equation (\ref{17}) can immediately be integrated to give
\begin{eqnarray}
e^{-\nu(r)}=1-\frac{C_1}{r}-\alpha^2r^2-2\alpha\beta
r-\beta^2,\label{19}
\end{eqnarray}
where $C_1$ is an integration constant. Substitution  of
$e^{-\nu(r)}$ into equation (\ref{18}) leads to
\begin{eqnarray}
e^{\mu(r)}=\frac{f(r)}{4r}\left(-C_2+2\alpha c
\int\frac{r^{5/2}dr}{f(r)^{3/2}}+2\beta
c\int\frac{r^{3/2}dr}{f(r)^{3/2}}\right)^{2},\label{20}
\end{eqnarray}
where $C_2$ is an integration constant and
\begin{eqnarray}
f(r)=-r+C_1+\alpha^2r^3+2\alpha\beta r^2+\beta^2r.
\end{eqnarray}
Although we have considered all the necessary equations/relations
(even with special choices), there are still a number of arbitrary
constants which do not let us find the unique vacuum solution of
the gravitational field equations on the brane because the
Birkhoff theorem does not apply here \cite{22,mak}. All the way,
among various possible solutions that depend on different choices
of arbitrary constants $c, \alpha, \beta$, we consider the two
following interesting cases. The choice $ c=0$ and use of equation
(\ref{20}) result in
\begin{eqnarray}
e^{\mu(r)}=e^{-\nu(r)}=1-\frac{C_1}{r}-\alpha^2r^2-2\alpha\beta
r-\beta^2.\label{M}
\end{eqnarray}
For small distances associated with the standard
stellar/astrophysical scales (e.g. solar system scales), the
corresponding metric for this case is the familiar Schwarzschild
solution. On larger distance scales associated with galaxies,
assuming a small value of $\alpha$ in order to neglecting its
second order term, the model differs from Einstein theory through
the following potential function
\begin{eqnarray}
\phi(r)=\frac{C_1}{2r}+\alpha\beta r+\frac{\beta^2}{2},\label{222}
\end{eqnarray}
which can be used to explain the galactic rotation curves
corresponding to the well-known dark matter problem without
resorting to dark matter and even without assuming any
new/modified mechanics (e.g. modified Newtonian dynamics [31]).

A second class of solutions of the system of equations
(\ref{17})-(\ref{new18}) can be obtained by considering
$c=\beta=0$ and $\alpha\neq0$. In this case by means of equation
(\ref{20}) we find
\begin{eqnarray}
e^{\mu(r)}=e^{-\nu(r)}=1-\frac{C_1}{r}-\alpha^2r^2.
\end{eqnarray}
The corresponding line element, choosing $C_1 =2GM=2m$ based on
proper correspondence principle in the limit of standard gravity,
takes the form
\begin{eqnarray}
ds^2=-\left(1-\frac{2m}{r}-\alpha^2r^2\right)dt^2+\frac{dr^2}
{\left(1-\frac{2m}{r}-\alpha^2r^2\right)}
+r^2\left(d\theta^2+\sin\theta^2d\varphi^2\right).\label{21}
\end{eqnarray}
Comparing the above result with the following line element for the
black hole solution in an asymptotically de Sitter space
\begin{eqnarray}
ds^2=-\left(1-\frac{2m}{r}-\frac{\Lambda}{3}r^2\right)dt^2+\frac{dr^2}{\left(1-\frac{2m}{r}-
\frac{\Lambda}{3}r^2\right)}
+r^2\left(d\theta^2+\sin\theta^2d\varphi^2\right),\label{21new}
\end{eqnarray}
the cosmological constant is found as $\Lambda=3\alpha^2$. This
positive value is in agreement with present observations. Now, let
us find the Kretschmann scalar of the metric (\ref{M}) as
\begin{eqnarray}
R_{\alpha\beta\gamma\delta}R^{\alpha\beta\gamma\delta}&=&\frac{4(6\alpha^4r^6+12\alpha^3\beta
r^5+10\alpha^2\beta^2r^4+4\alpha\beta^3r^3+\beta^4r^2+4m\beta^2r+12m^2)}{r^6},\label{23}
\end{eqnarray}
the singularity at $r=0$ is an intrinsic singularity. In the case
that $\alpha=0$ we have the Schwarzschild horizon $r=2m$, and for
$m=0$ we have the de Sitter horizon $r=\frac{1}{\alpha}$. For
$\sqrt{27}m<\frac{1}{\alpha}$ there are two horizons
\begin{eqnarray}\label{root1}
r_1=\frac{2}{\sqrt{3}\alpha}\cos\frac{\Theta}{3},
\end{eqnarray}
\begin{eqnarray}\label{root2}
r_2=-\frac{1}{\sqrt{3}\alpha}\left(\cos\frac{\Theta}{3}-\sqrt 3
\sin\frac{\Theta}{3}\right),
\end{eqnarray}
where $\Theta$ is given by
\begin{equation}
\label{d} \cos\Theta=-\sqrt{27}m\alpha.
\end{equation}
If $m=0$ we have $r_2=0$ and $r_1=\frac{1}{\alpha}$, then we call
$r_1$ the cosmological horizon generalized when $m\neq 0$, and
$r_2$ the black hole horizon generalized when $\alpha\neq 0$. For
$\sqrt{27}m=\frac{1}{\alpha}$, $r_1$ and $r_2$ coincide and there
is only one horizon,
\begin{equation}
\label{mass} r=\frac{1}{\sqrt{3}\alpha}.
\end{equation}
In general, the range for $r_1$ and $r_2$ is given by
\begin{equation}
\label{leq} 0 \leq r_2 \leq \frac{1}{\sqrt{3}\alpha} \leq r_1 \leq
{\frac{1}{\alpha}}.
\end{equation}
For $\sqrt{27}m>\frac{1}{\alpha}$ there are no horizons.

Finally, we should mention that if we had worked as in the usual
brane-world models where the Israel junction condition is used to
calculate the extrinsic curvature in terms of the energy-momentum
tensor on the brane and its trace, that is
\begin{eqnarray}
K_{\mu\nu}=-\frac{1}{2}\alpha^{* 2}(\tau_{\mu\nu}-\frac{1}{3}\tau
g_{\mu\nu}),\label{new55}
\end{eqnarray}
where $\alpha^{*}$ is proportional to the gravitational constant
in the bulk, then by substituting equation (\ref{new55}) into
equation (\ref{new5}), the vacuum field equation (\ref{A9}) with a
constant curvature bulk would have been reduced to the vacuum
field equations in the standard general relativity $G_{\mu\nu}=0$;
but, the $Q_{\mu\nu}$ term here modifies the vacuum field
equations on
 the brane and this is because it originates from the so-called
 confining potential
and not the usual junction condition.
\section{Conclusion}
In this paper, we have considered the vacuum field equations in a
brane world model where the matter is confined to the brane
through the action of a confining potential, rendering the use of
any junction condition redundant. We have obtained the exact
solutions for static black holes localized on a 3-brane in a
constant curvature bulk. A particular solution of the field
equations represents a Schwarzschild-de Sitter black hole in the
presence of a positive cosmological constant. This shows that the
extra term $Q_{\mu\nu}$ may play the role of dark energy,
supporting our previous results that the accelerated expansion of
the universe could be explained in a purely geometrical fashion
based on the extrinsic curvature. Another interesting solution to
the model considered here is one in which we can have a proper
potential to explain the galaxy rotation curves without assuming
the existence of dark matter and without working with new modified
theories (modified Newtonian dynamics [31,32]).

\end{document}